%% file: llmtop.tex
\documentclass[aps,prd,twocolumn,showpacs,superscriptaddress]{revtex4-1} % for collaboration review and submission
\usepackage{lineno}
\usepackage{setspace}
\usepackage{graphicx} % needed for figures
\usepackage{dcolumn} % needed for some tables
\usepackage{bm}  % for math
\usepackage{amssymb} % for math
\usepackage{amsmath}
\usepackage{setspace}

\newcommand{\dzero}  {D0}
\newcommand{\ttbar}  {\mbox{$t\bar{t}$}}
\newcommand{\qqbar}  {\mbox{$q\bar{q}$}}
\newcommand{\ppbar}  {\mbox{$p\bar{p}$}}
\newcommand{\met}  {\mbox{$\not\!\!E_T$}}
\newcommand{\rar}  {\rightarrow}
\newcommand{\mt}	{\mbox{$m_{t}$}}
\newcommand{\herwig} {{\sc herwig}}
\newcommand{\pythia} {{\sc pythia}}
\newcommand{\alpgen} {{\sc alpgen}}

\newcommand{\mcnlo}	{{\sc mc@nlo}}
\newcommand{\geant}  {{\sc geant}}

\newcommand{\metcalc}	{\mbox{$\not\!\!E_{T}^{\rm \/ calc}$}}
\newcommand{\metobs}	{\mbox{$\not\!\!E_{T}^{\rm \/ obs}$}}
\newcommand{\metunc}	{\mbox{$\not\!\!E_{T}^{u}$}}

\newcommand{\pt}  {$p_T$}

\newcommand{\mtfit}	{$m_t^{\rm fit}$}
\newcommand{\mtmc}	{$m_t^{\rm MC}$}
\newcommand{\etal}	{\textit{et al.}}
\newcommand{\ljets}	{$\ell+$jets}
\newcommand{\gamjet}	{$\gamma+$jet}

\begin{document}

%%% the following information is for internal review, please remove them for submission

% the following line is for submission 
\hspace{5.2in} \mbox{FERMILAB-PUB-15-338-E}

\title{Precise measurement of the top quark mass in dilepton decays using optimized neutrino weighting}
\input author_list.tex % input Dzero author list

\date{\today}

%---------------------------------------------------------------------------------------------------

%\linenumbers

\begin{abstract}
We measure the top quark mass in dilepton final states of \ttbar\
events in \ppbar\ collisions at $\sqrt{s}=1.96$~TeV,
using data corresponding to an integrated luminosity of $9.7$~fb$^{-1}$
at the Fermilab Tevatron Collider.
The analysis features a comprehensive optimization of the neutrino weighting method to minimize the statistical uncertainties.
We also improve the calibration of jet energies using the calibration
determined in \ttbar\ $\rar$ lepton+jets events, which reduces the otherwise limiting
systematic uncertainty from the jet energy scale.
The measured top quark mass is $\mt=173.32\pm1.36({\rm stat})\pm0.85({\rm syst})$~GeV.
\end{abstract}

\maketitle
\vskip 1.0in 

\section{Introduction}

The discovery of the top quark in 1995~\cite{tobsD0,tobsCDF} completed the three quark families of the standard model (SM).
Since then, the top quark has been one of the focal points of the Fermilab Tevatron and of the CERN LHC programs.
The top quark stands out because of its large mass, \mt, which is a fundamental parameter in the SM.
Its Yukawa coupling to the Higgs boson, $Y_t=\sqrt{2} m_t/v$, where $v$ is the
vacuum expectation value of the Higgs field, is close to unity, implying that the top quark may play a special role in electroweak symmetry breaking.
In addition, \mt\ is linked to the $W$ and Higgs boson masses, $M_W$ and $M_H$, through radiative corrections~\cite{GfitterSM}.
Following the Higgs boson discovery~\cite{HobsATLAS,HobsCMS}, a precise measurement of \mt\
provides a test of the electroweak sector of the SM
and information on whether our universe resides 
in a stable or metastable region of that theory~\cite{ewstab1,ewstab2,ewstab3}.  
The short lifetime of the top quark prevents its
confinement in the strong color field, since top quarks decay before hadronizing.
This allows a particularly precise study of pure quantum chromodynamic (QCD) effects.
A comparison of the measured \mt\ and the \mt\ extracted from cross section measurements~\cite{mtpoled0,mtpoleCDF,mtpoleCMS,mtpoleATLAS} 
may provide a probe of higher order and soft QCD corrections to the observed mass~\cite{mtpole}.

Assuming the SM branching ratio of $t\rar Wb\approx 100$\%, \ttbar\ decays yield 
distinct final state categories according to the number of charged leptons
with high transverse momentum (\pt) from $W$ boson decays.
Dilepton ($2\ell$, $\ell=e$ or $\mu$) events, such as $ee$, $e\mu$, and $\mu\mu$, with neutrinos from two $W\rightarrow \ell\nu$ decays, are relatively rare but have low background.
We present a measurement of \mt\ using \ppbar\ collider data collected with the \dzero\
detector at the Fermilab Tevatron collider, corresponding to an integrated luminosity
of 9.7~fb$^{-1}$, in events with two high-\pt\ electrons or muons of opposite electric charge.
Two high-\pt\ jets must also be observed, one of which must be identified as being 
consistent with originating from a $b$ quark. 
This analysis is based on our previous dilepton measurement~\cite{llmtop12}, but with increased integrated luminosity and multiple optimizations to improve the precision of \mt.
We reduce the dominant statistical contribution to the uncertainty on \mt\
through an optimization of the methods for kinematic reconstruction and statistical analysis. 
Lacking a dijet signature from $W\rightarrow q\bar{q}'$, which is present in \ttbar\ $\rar$ lepton+jets (\ljets) events
and was used to improve the precision of jet energy calibration with a $W$ mass constraint~\cite{ljets14}, previous dilepton analyses at the Tevatron have
reached a sensitivity limit imposed by standard jet calibration methods~\cite{llmtopME,cdfllmtop}.
Progress in calibrating jet energies in the dilepton channel~\cite{llmtop12} provides improved cross-checks across different channels and a more significant contribution from the dilepton channel to the world average \mt~\cite{worldavg}.
For comparison, the most recent measurements of \mt\ in the dilepton channel from CDF, ATLAS, and CMS are, respectively,
$\mt=171.5\pm1.9({\rm stat})\pm2.5({\rm syst})$~GeV~\cite{CDF2l},
$\mt=173.79\pm0.54({\rm stat})\pm1.30({\rm syst})$~GeV~\cite{ATLAS2l}, and
$\mt=172.50\pm0.43({\rm stat})\pm1.46({\rm syst})$~GeV~\cite{CMS2l}.
In this analysis, we substantially reduce the otherwise dominant uncertainty in the jet energy scale
by applying the methods of Ref.~\cite{llmtop12}.  

\section{Detector and Data Sample}

\subsection{Detector}

The \dzero\ detector~\cite{d0nim,2bsmt} has a central-tracking system, consisting
of a silicon microstrip tracker and a central
fiber tracker, both located within a 1.9 T superconducting
solenoidal magnet, with designs optimized for
identification of the \ppbar\ collision vertex and track reconstruction at
pseudorapidities~\cite{pseudorapidity} of $|\eta| <$ 3 and
$|\eta| <$ 2.5, respectively. The liquid-argon/uranium calorimeter
has a central section covering $|\eta|$ $\leq$ 1.1, and two end sections that extend
coverage to $|\eta|$ $\approx$ 4.2, with all three housed in separate
cryostats. An outer muon system, covering $|\eta| <$ 2,
consists of a layer of tracking detectors and scintillation
trigger counters in front of 1.8 T iron toroids, followed by
two similar layers after the toroids.
 
\subsection{Object Reconstruction}

We require electrons to satisfy an identification criterion based on 
boosted decision trees~\cite{eleNIM} using calorimeter and tracking information.
Muons must satisfy requirements that match hits in the muon system
to a track in the central tracking detector
that is required to have a small distance of closest approach
to the beam axis~\cite{muonNIM}. We require 
hits in the muon layers inside and outside the toroid. 
Muons and charged hadron momenta are measured in the central tracking detector,
while electron, photon ($\gamma$), jet, and charged hadron energies are measured in the calorimeters.
Muons must be isolated from jets and from nearby tracks.
Electrons and muons must have their extrapolated track trajectories
isolated from calorimeter energy depositions greater than an energy threshold.  
Electrons and muons must have $p_T$ $>$ 15 GeV, and $|\eta|<2.5$ and $<2.0$, respectively.
We reconstruct jets using an iterative, midpoint-seeded cone algorithm with a cone parameter of
${\cal R}_{\rm cone}=0.5$~\cite{jetalgo}. 
Jets with embedded muons from the decay of $b$-hadrons require an additional
correction to jet energy to account for the associated neutrino.
A multivariate discriminant~\cite{bjetid} is used to identify jets
that contain a $b$-hadron (i.e., $b$ jets) from a vertex displaced from the interaction point.
We define the missing transverse momentum (\met)
attributed to the escaping neutrinos as the negative of the
vector sum of all transverse components of calorimeter cell energies,
corrected for the measured muon momenta and the response of the
calorimeter to electrons.  We also correct \met\ for detector response in the
jet energy calibration, as described below.
Details of object reconstruction are provided in Ref.~\cite{llafb13}.

\subsection{Standard Jet Energy Calibration}

We calibrate the energy of jets to be the energy of the particle jets reconstructed using 
the midpoint algorithm~\cite{jetalgo}. We correct for the effects of 
the calorimeter response to particle constituents of jets,
energy leaking into the cone from particles directed outside it,
as well as energy deposits outside the cone from particles inside it~\cite{jetscale}. 
Charged hadrons have an energy-dependent 
response that is lower than that of electrons and photons.
We therefore apply corrections obtained from \gamjet\ events to
account for the energy dependence of the jet response in the central $|\eta|$ region. 
We also apply a relative $\eta$-dependent correction 
obtained from \gamjet\ and dijet events. 
We employ the same methods to calibrate jet energies independently in the Monte Carlo (MC) simulation
and in data. The MC is used to help study potential biases in the data.  We incorporate a
correction for jets in the MC simulation that accounts for the difference in single-particle response between data and MC.  This procedure ensures that the flavor dependence of the 
jet response in data is replicated in MC.
In the MC we account for multiple \ppbar\ interactions
by correcting the jet energy to the particle level of 
only those particles that are directed within the jet cone at particle level.
The typical systematic uncertainty in the energy calibration of each jet in the
dilepton sample is 2\%.  This precision is limited by systematic uncertainties of the $\gamma+$jet method in the 
$p_T$ range of jets in \ttbar\ events. Details about this ``standard jet energy scale'' calibration can be found in 
Ref.~\cite{jetscale}.
We require that jets have $p_T>20$~GeV and $|\eta|<2.5$ after calibration, but before applying additional
corrections from the $W\rightarrow q\bar{q}'$ constraint in the \ljets\ channel discussed below. 

\section{Absolute Jet Calibration from a $W\rightarrow q\bar{q}'$ Constraint}

As in Ref.~\cite{llmtop12}, we apply a multiplicative correction factor to the energy 
of jets in data based on an analysis of \ttbar$\rightarrow$\ljets\ events 
using the $W\rightarrow q\bar{q}'$ decays as a constraint.  
Application of this factor, $1.0250\pm0.0046\,({\rm stat})$~\cite{ljets14}, improves the agreement 
between MC and data and allows us to use its uncertainty to 
reduce the uncertainty on the absolute energy scale by a factor of $\approx$~4 relative to 
the standard jet energy scale, while retaining its $\eta$ and $p_T$ dependence.  
To apply this scale, which comes from light-quark jets, to the dilepton sample, which has $b$ jets,
it is important to ensure that the variation in the ratio of data over MC jet response between different 
flavors be placed on an equal footing. The standard jet energy scale~\cite{jetscale} achieves this on a 
jet-by-jet basis by using single particles in MC jets to correct. This ensures that the energies of $b$ jets
in dilepton simulated samples agree with those of $b$ jets in the dilepton data sample at the same level as light-quark jets.
Aside from fragmentation differences between data and MC which are
discussed below, this approach justifies the use of the \ljets\ constraint in the dilepton channel.

\section{Event Selection}

The \ttbar\ candidate events in the $ee$ and $\mu\mu$ channels are required to pass single-lepton triggers.
The full suite of triggers is used for selecting $e\mu$ events.
The dilepton event selection before optimization is described in Ref.~\cite{llafb13}.
We optimize the selection based on MC events to provide the smallest expected
statistical uncertainty in \mt. 
We require two isolated leptons with opposite electric charge.
We require at least two jets, where at least one of the two jets with highest \pt\ must 
be identified as a $b$ jet. 
For the $e\mu$ channel, our selections have an efficiency for tagging $b$ jets
of 72\%, and a light-quark mistag rate of 12\% in the central region in $\eta$.
The same-flavor channels employ slightly tighter $b$ tagging requirements and thus
have a few percent lower efficiency, and 30\% lower mistag rate.
We require events in the $\mu\mu$ channel to have \met$>40$~GeV.  This \met\ selection is also
applied to $ee$ events when the dielectron invariant mass is between 70 and 100~GeV, to reduce
the $Z\rightarrow ee$ background contribution.
We define a \met\ significance variable, $\cal S$, which measures the likelihood for the observed \met\ to be a fluctuation from $\met=0$~GeV.
We require $\cal S$ $>3.5$~(4) for the $ee$ ($\mu\mu$) channel.
We require $e\mu$ events to have $H_T>100$~GeV, where $H_T$ is the scalar sum of the 
$p_T$ of the two highest-$p_T$ jets and of the lepton with highest $p_T$.
The $H_T$, $b$ tagging, and \met-based requirements are optimized
to minimize the expected statistical uncertainty on \mt\ in each channel. 
The expected signal-to-background (S/B) ratio is $\approx$~7 for these channels.
These requirements yield a 3\% improvement in
statistical precision in \mt\ relative to the selections in Ref.~\cite{llmtop12}.
After implementing all these selections, we obtain 340, 115, and 110 events in the $e\mu$,
$ee$ and $\mu\mu$ channels, respectively. 

\section{Modeling Signal and Background}

The \ttbar\ events are simulated at 15 mass points over the range $130 \le m_t^{\rm MC} \le 200$~GeV
using the tree level generator \alpgen\ 2.11~\cite{alpgen} with up to 2 additional light partons and
\pythia\ 6.409~\cite{pythia} with modified underlying event Tune A for parton showering and hadronization. Here, 
\mtmc\ refers to the input mass in \alpgen.  An additional,
larger sample is generated at \mtmc$=172.5$~GeV to study systematic uncertainties. 
We normalize the \ttbar\ production cross section to $\sigma_{t\bar{t}}=7.24\pm0.23$~pb~\cite{nnloxscalc}, which is calculated at 
next-to-next-to-leading order with a
next-to-next-to-leading logarithm soft gluon resummation.
The main backgrounds arise from three sources:
$Z/\gamma^{*}\rightarrow \ell^{+}\ell^{-}$,
diboson ($WW$, $WZ$, and $ZZ$) processes, and instrumental effects. 
We model the $Z/\gamma^{*}$ background using \alpgen\ with up to 2 light partons and 
\pythia\ for showering and hadronization. We employ \pythia\ for the diboson background. 
The instrumental background arises from $W+$jets, multijet, or \ljets\ \ttbar\ events where
one or two jets are either mis-identified as electrons, or they contain a hadron decaying
to a non-isolated lepton that passes
our selection. This background is estimated from data as in Ref.~\cite{llafb13}.
We apply a full detector simulation based on \geant\ 3.14~\cite{geant} for all simulated events.
The objects reconstructed in simulation are smeared to ensure that their resolutions reflect those in data.
Scale factors in object efficiencies are applied to improve agreement between data and MC.

\section{Kinematic Reconstruction}

\subsection{Neutrino Weighting}

The presence of two neutrinos in the \ttbar\ decay
makes it impossible to fully constrain the kinematics and thus extract a
unique \mt\ measurement from each event.
Given the measured momenta of leptons, jets and \met, the available
constraints from $M_W$, and the condition $m_t=m_{\bar{t}}$, we are missing one
constraint to provide full \ttbar\ reconstruction in dilepton events. We integrate
over the phase space of neutrino rapidities for chosen values of hypothesized \mt\ ($m_t^{h}$)~\cite{nuWTR1},
and compare $\metcalc$, the vector sum of neutrino momentum solutions at each chosen point of phase space, to the observed $\metobs$ to determine a
``weight'' $\omega$ characterizing the level of agreement:
\begin{equation}
\label{eq:weight}
\omega=\frac{1}{N}\sum_{i=1}^{N}\prod_{j=x,y} \exp\left(- \frac{({\not\!\!E_{T_{j,i}}^{\rm \/ calc}}-{\not\!\!E_{T_j}^{\rm \/ obs}})^2}{2\sigma^2_{{\not\!E_{T}^{u}}_{j}}}\right),
\end{equation}
\noindent where $i$ runs over all neutrino solutions for any two possible jet-lepton assignments in the
\ttbar\ final state (up to $N = 8$), $j$ stands for the two orthogonal coordinates in the transverse plane ($x$ and $y$),
and $\sigma_{{\not\!E_{T}^{u}}_{j}}$ is the resolution of the $x$ or $y$ component of unclustered \met.
The unclustered \met\ (\metunc)
is the magnitude of the vector sum of all energy depositions in the calorimeter that were not
included in lepton or jet reconstruction. 
The quantity $\sigma_{\not\!E_{T}^{u}}$ is parameterized to describe the difference between the neutrino 
momentum solutions and the measured \met\ in the weight calculation.
We perform this calculation over a range of $m_t^{h}$, integrating $\omega$ over the neutrino
phase space, to yield a distribution of $\omega(m_t)$ versus $m_t^{h}$.
Prior studies~\cite{llmtop09} have shown that the first two moments ($\mu_{\omega}, \sigma_{\omega}$) of this 
distribution extract most of the information about \mt.
The analysis of Ref.~\cite{llmtop12} used the range of $m_t^{h}$ values between 80 and 330~GeV in 1 GeV step and a \metunc\ resolution of 7~GeV in the weight calculation.
The new optimized determination of these parameters is briefly summarized below.

\subsection{Optimization of Weight Calculation Parameters}

After applying the methods described above to improve the jet energy calibration,
the statistical contribution is the dominant source of measurement uncertainty on \mt\ in the dilepton channel.
We therefore examine the parameters used for the kinematic
reconstruction of \ttbar\ events and for the maximum likelihood fit to
reduce the expected statistical uncertainty. At each step, we verify
through MC simulations that the optimization
does not increase the systematic uncertainty.

All neutrino solutions and jet assignments 
yield mass estimators such as $\mu_{\omega}$ that are correlated with \mt.
However, the correlation is substantially greater, and $\mu_{\omega}$ values are less biased,
when the correct jet assignments and solutions of neutrino momenta are chosen. 
Since now \mt\ has been measured with high precision~\cite{worldavg}, we can optimize the range of $m_t^{h}$ based on known values of \mt.
Considering a wide range in $m_t^{h}$ causes incorrect configurations to overwhelm the correct
configuration, thereby worsening the mass resolution. Likewise, scanning 
over too narrow a range biases the background and worsens the mass sensitivity by causing
\ttbar\ and background distributions to be similar.
Examination of a two-dimensional grid of upper and lower limits of the mass range yields
the optimal range of $m_t^{h}=115$ to 220~GeV in 1 GeV steps.
The value of $\sigma_{{\not\!E_{T}^{u}}_{j}}$ also has a noticeable impact on the expected precision of
the analysis.
Properly accounting for the full difference between the calculated and measured \met, 
including effects such as mismatches between the true neutrino $p_T$ and that taken from finite binning of the 
neutrino rapidity distributions, or the difference in the number of jets used to derive the calculated and measured \met,
requires that we optimize the value of the \met\ resolution parameter.  
We scan $\sigma_{{\not\!E_{T}^{u}}_{j}}$ for a wide range from 7 to 100~GeV and find 25~GeV is optimal.
Combined, these optimizations improve the expected combined statistical uncertainty on
\mt\ by 11\% compared to the parameters in Ref~\cite{llmtop12}.

\subsection{Efficiency of Kinematic Reconstruction and Event Yields}

\begin{figure}
\includegraphics[width=8cm]{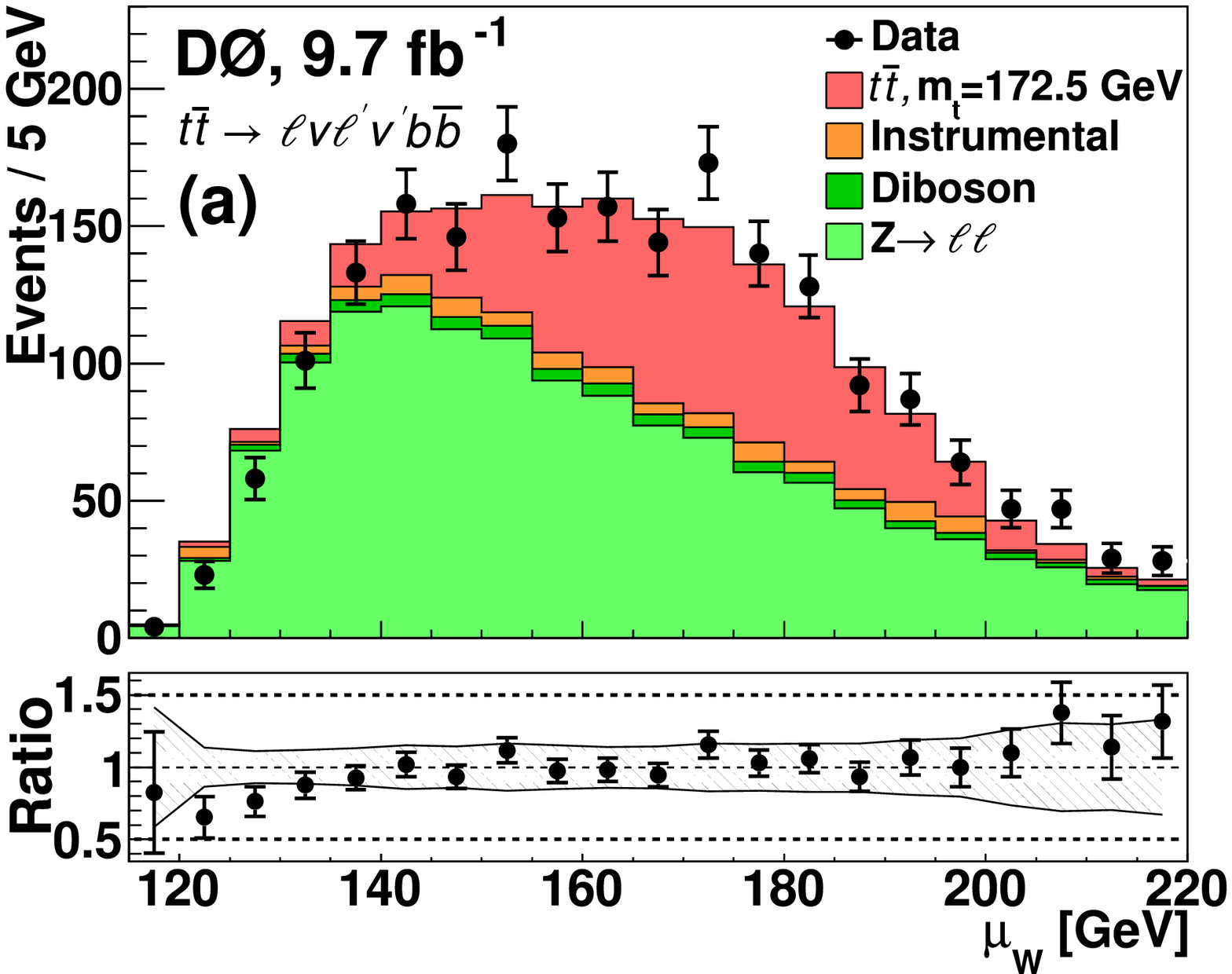}
\includegraphics[width=8cm]{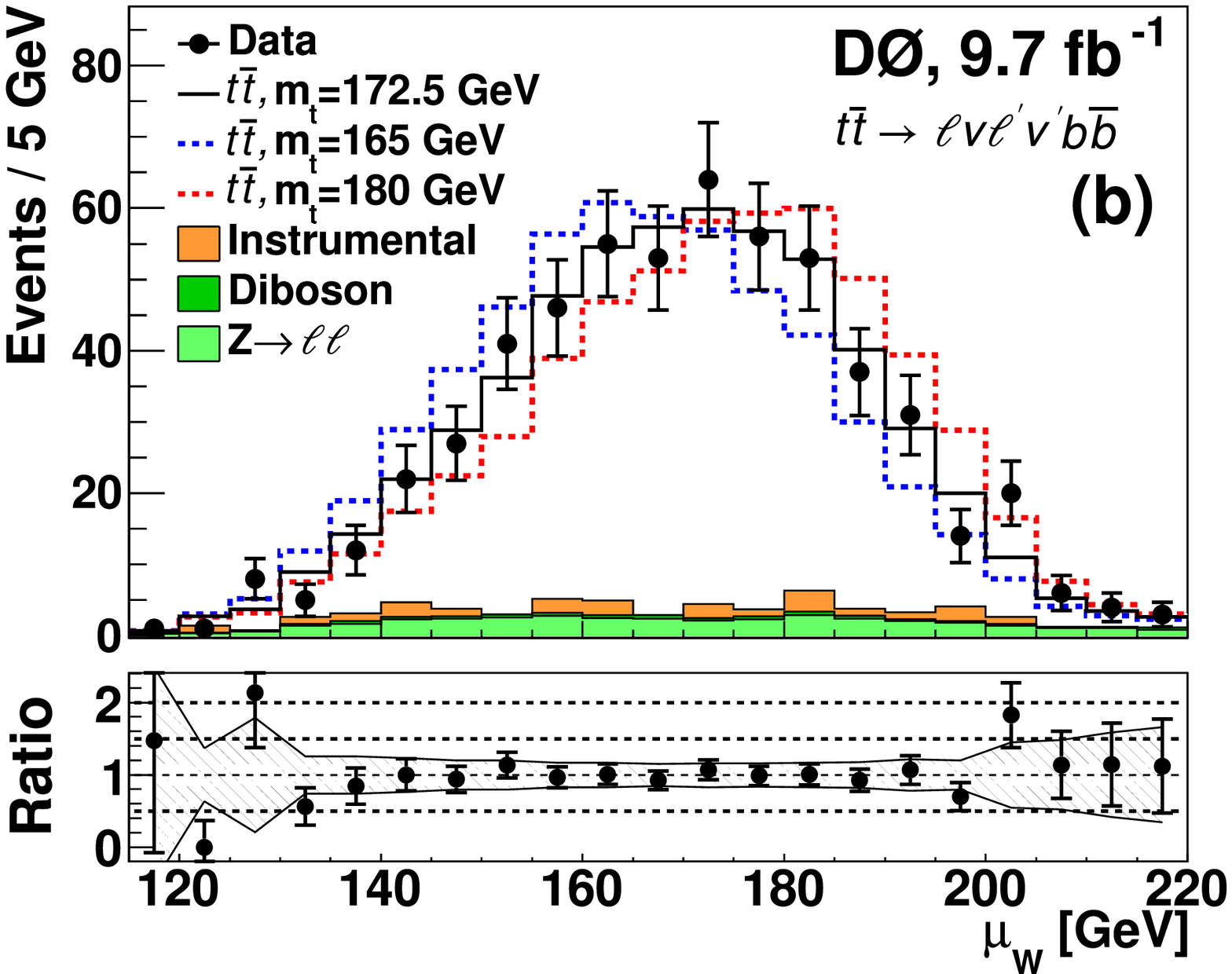}
\caption{\label{fig:means}{The distribution in the mass estimator, $\mu_w$, for the combination of the
$ee$, $e\mu$, and $\mu\mu$ channels for (a) the preselected sample and (b) the final event sample.
The MC events are normalized separately to the number of observed events in data in each channel.
The ratios show the total number of observed events divided by the number of expected events in a given bin of $\mu_w$ for \mtmc$= 172.5$~GeV.
The band of systematic uncertainty is shown as the shaded area in the ratio plots, which includes contributions from the dominant sources: jet energy scale,
lepton identification, lepton momentum scale, luminosity, $b$ quark modeling, initial and final state radiation,
color reconnection, as well as hadronization and higher-order QCD effects for \ttbar\ events.}}
\end{figure}

Events used in the analysis must have at least one pair of neutrino solutions for at least one $m_t^{h}$ value.
The efficiency for this kinematic reconstruction is over 99\% for \ttbar\ events, 
and 91\% to 98\% for the background. In the final sample, a total of
336, 113, and 109 events in the $e\mu$, $ee$, and $\mu\mu$ channels, respectively, pass the kinematic reconstruction.
The expected sum of \ttbar\ and background yields and their corresponding 
asymmetric total uncertainties (stat$\oplus$syst) are $298.1\substack{+22.1 \\ -27.2}$,
$106.5\substack{+10.4 \\ -11.6}$, and $103.5\substack{+7.4 \\ -9.1}$ events
for the $e\mu$, $ee$, and $\mu\mu$ channels, respectively.
The distributions of the mass estimator $\mu_{\omega}$ in a preselected sample, omitting requirements
on $b$ tagging, \met, \met\ significance, and $H_T$, are shown in Fig.~\ref{fig:means}(a).
The \ttbar\ component is evident in the preselected data.  
The mass dependence of the $\mu_{\omega}$ 
distribution is given in Fig~\ref{fig:means}(b) for three \mtmc\ mass points with all selections applied.

\section{Extracting the Top Quark Mass}

\subsection{Maximum Likelihood}

We perform a binned maximum likelihood fit to the extracted
moment distributions $[\mu_{\omega},\sigma_{\omega}]$ in data. Expected 
probability densities are calculated using the MC samples for each of the 16 \mt\ points, yielding a
two-dimensional probability density $h_{\rm S}(\mu_{\omega},\sigma_{\omega}|m_t^{\rm MC})$ distribution parametrized by \mt.
Background samples are used to construct a background template for each channel,
$h_{\rm B}(\mu_{\omega},\sigma_{\omega})$, with each background contributing according to its expected
yield.  Bins in signal templates with no events are given a weighted value corresponding to a single signal MC event to ensure that the log of likelihood is not infinite. 
The likelihood is given by:
\begin{eqnarray}
\label{eq:likeli}
\lefteqn{\mathcal L({\mu_{\omega}}_{\{1..N\}},{\sigma_{\omega}}_{\{1..N\}},N \mid n_{\rm S}, n_{\rm B}, m_t) = } \nonumber \\
& & \prod_{i=1}^N\frac{n_{\rm S} \cdot h_{\rm S}({\mu_{\omega}}_i, {\sigma_{\omega}}_i \mid
m_t) + n_{\rm B} \cdot h_{\rm B}({\mu_{\omega}}_i, {\sigma_{\omega}}_i)}{n_{\rm S} + n_{\rm B}},
\end{eqnarray}
where $N$ is the number of observed events in data, $n_{\rm S}$ is the expected number of \ttbar\ events 
(for $m_t=172.5$~GeV), and $n_{\rm B}$ is the expected total number of background events.
We fit ($-\ln{\mathcal L}$) versus \mtmc\ to a parabola in a window of \mtmc\ that is iteratively varied until a stable minimum is found.
We take the minimum of the final parabola to be the fitted top quark mass, \mtfit.
The uncertainty on the fitted mass is obtained by considering the \mtmc\ range over which
the fit function increases by 0.5 units in ($-\ln{\mathcal L}$) above this minimum.
Using pseudo-experiments, we optimize the template binning of each channel separately 
in a two-dimensional grid that lets $\mu_{\omega}$ and $\sigma_{\omega}$
bin sizes vary independently. Finer binning in $\mu_{\omega}$ and $\sigma_{\omega}$, especially for the $e\mu$ channel, improves the
expected statistical precision in \mtfit\ by 5\%.
The fitted mass window is optimized to $\pm 15$~GeV for all channels.
Taking all the optimizations together, 
including event selection, weight calculation, and maximum likelihood fitting,
the statistical sensitivity of this analysis is improved relative to Ref.~\cite{llmtop12} by 20\%
beyond the 35\% gain expected from increased integrated luminosity. 

\subsection{Ensemble Testing and Data Results}

\begin{table}
\caption{\label{tab:calib} 
Slopes, offsets, and pull widths
of the \mt\ calibration and the expected statistical uncertainties in the mass ($\sigma_{\it{m}_{t}}$) for the $ee$,
$e\mu$, and $\mu\mu$ channels, and their combination.}
\begin{tabular}{l|c|c|c|c}
\hline
\hline
 	& Slope 	& Offset [GeV] & Pull width & $\sigma_{\it{m}_{t}}$~[GeV] \\
\hline
$ee$		& $0.984\pm0.004 $ & $0.671\pm0.043 $ & $ 0.994 $ & 2.98\\
$e\mu$		& $0.986\pm0.006 $ & $0.548\pm0.065 $ & $ 0.998 $ & 1.72\\
$\mu\mu$	& $0.989\pm0.010 $ & $0.717\pm0.103 $ & $ 1.004 $ & 3.31\\
$2\ell$ & $0.988\pm0.006 $ & $0.617\pm0.063 $ & $ 0.995 $ & 1.35\\
\hline
\hline
\end{tabular}
\end{table}

We obtain a linear relationship between \mtfit\ and \mtmc\ by performing randomized
pseudo-experiments using all signal mass points.
The numbers of signal and background events in the pseudo-experiments are allowed to fluctuate
within their Poisson uncertainties around their expected values.  We require that the 
total number of events matches that observed in data.
To minimize the effect of statistical fluctuations on our systematic uncertainties,
we optimize the number of pseudo-experiments by dividing the MC sample into five 
subsamples, and measure systematic uncertainties with each subsample.  We calculate
the RMS of the five uncertainties, average over all systematic effects, 
and divide by $\sqrt{5}$ to
estimate the statistical component of systematic uncertainties.  The 
average RMS decreases until we oversample, or reuse, the \ttbar\ MC events 
by roughly a factor of three.  This corresponds to 3000 pseudo-experiments.
We perform a linear fit of \mtfit\ versus \mtmc\ to obtain a calibration slope and offset for \mtfit\ using 3000 pseudo-experiments:
\begin{equation}
\label{eq:calibration}
  \it{m}_{\it{t}}^{\rm fit} = {\rm Slope} \cdot ( \it{m}_{\it{t}}^{\rm MC} - {\rm 170} ) + {\rm Offset} + {\rm 170}.
\end{equation}
We account for oversampling by increasing the statistical 
uncertainties at each mass point by the appropriate oversampling factor.  
Likewise, we compute the pull, or the ratio of \mtfit$-$\mtmc\ over the average
estimated uncertainty at each mass point. 
The slopes of \mtfit\ versus \mtmc\ are close to 1,
and pull widths are consistent with unity, as shown in Table~\ref{tab:calib}. 
We calculate the final \mt\ by correcting \mtfit\ from a given measurement by
the slope and offset.  We correct
the statistical uncertainty using the slope and the pull width.
The expected corrected statistical uncertainties
for each channel are given in Table~\ref{tab:calib}. 
In data, we obtain corrected, fitted
\mt\ values of $\mt=171.86\pm1.71({\rm stat})$, $173.99\pm3.04({\rm stat})$, and
$178.58\pm3.56({\rm stat})$~GeV for the $e\mu$, $ee$, and $\mu\mu$ channels respectively, 
and $\mt=173.32\pm1.36({\rm stat})$~GeV for the combined channels. 

\section{Systematic Uncertainties}

Systematic uncertainties summarized in Table~\ref{tab:syst} arise from jet energy calibration, object
reconstruction, modeling of \ttbar\ and background events, and the mass-extraction method.
The energies of jets are shifted up and down by 
the uncertainty on the absolute
energy scale, which is taken from \ljets\ events, thereby providing shifts in
\mt.  This scale is appropriate for
light-quark jets, which, after correcting for jet flavors to improve the agreement between data and MC,
have different kinematic distributions than
$b$ jets from \ttbar\ decays.  
We calculate a residual uncertainty due to the kinematic differences between  
the \ljets\ calibration sample
and dilepton sample of $b$ jets. 
We use separate up and down estimates to extract the energy- and
$\eta$-dependent shifts in \mt\ based
on uncertainties in the standard jet energy scale relative to their average value in the \ljets\ calibration 
sample.  We cross-check this with an alternative method 
that applies shifted light-quark jet energy scales to
$b$ jets in the \ljets\ channel~\cite{ljets14}. 
These methods agree, and thereby validate the use of
the \ljets\ scale as a jet calibration.  We also cross-check using a jet-energy-dependent linear
parameterization of the residual jet energy scale as in 
Ref.~\cite{ljets14}, obtaining results that do not exceed our estimate 
of uncertainties from the jet energy scale.
To estimate the uncertainty corresponding to possible differences in the 
flavor dependence of the MC scale relative to data, we change the single-particle responses up and down 
by their uncertainties and obtain the shift in \mt.  To estimate the possible dependence on the
$b$ quark fragmentation in the MC, we replace the \pythia\ $b$ quark fragmentation
function with the Bowler scheme~\cite{bowler}, and compare \mt\ with the Bowler free
parameters tuned to LEP (ALEPH, OPAL, and DELPHI) or SLD data~\cite{LEPSLD}. 

\begin{table}
\caption{\label{tab:syst} 
Systematic uncertainties on \mt\ 
for the combined dilepton measurement using 9.7~fb$^{-1}$ of integrated luminosity. 
For symmetrized uncertainties, the ``$\pm$" symbol indicates that the corresponding systematic 
parameters in MC are positively correlated with \mt\ in data,
and the ``$\mp$" symbol indicates an anticorrelation.
The uncertainties shown as + or $-$ only are 
computed by comparing a standard choice with an alternate, but are symmetrized in 
calculating the total uncertainty.}
\begin{tabular}{l|c}
\hline
\hline
Source & $\sigma_{m_t}$ [GeV] \\
\hline
Jet energy calibration & \hspace{1ex} \\
\hskip 0.5cm Absolute scale   		& $ \mp 0.47 $\\
\hskip 0.5cm Flavor dependence			 & $ \mp 0.27 $\\
\hskip 0.5cm Residual scale			 & $\substack{+0.36 \\ -0.35}$\\
\hskip 0.5cm $b$ quark fragmentation		 & $ +0.10 $\\
\hline
Object reconstruction & \hspace{1ex} \\
\hskip 0.5cm Trigger			 & $  -0.06 $ \\
\hskip 0.5cm Electron $p_T$ resolution		& $ \pm 0.01 $ \\
\hskip 0.5cm Muon $p_T$ resolution		 & $ \mp 0.03 $ \\
\hskip 0.5cm Electron energy scale		& $ \pm 0.01 $ \\
\hskip 0.5cm Muon $p_T$ scale			& $ \pm 0.01 $ \\
\hskip 0.5cm Jet resolution			 & $ \mp 0.12 $ \\
\hskip 0.5cm Jet identification			 & $  +0.03 $ \\
\hskip 0.5cm $b$ tagging 			 & $  \mp 0.19 $ \\
\hline 
Signal modeling & \hspace{1ex} \\
\hskip 0.5cm Higher-order effects		 & $ -0.33 $\\
\hskip 0.5cm ISR/FSR				 & $ \pm 0.15 $\\
\hskip 0.5cm $p_T(\ttbar)$				& $ -0.07 $\\
\hskip 0.5cm Hadronization				 & $  -0.11 $\\
\hskip 0.5cm Color reconnection			 & $  -0.22$\\
\hskip 0.5cm Multiple $p\bar{p}$ interactions				& $ -0.06 $\\
\hskip 0.5cm PDF uncertainty			  & $\pm 0.08 $\\
\hline
Background modeling & \hspace{1ex} \\
\hskip 0.5cm Signal fraction			 & $ \pm 0.01 $\\
\hskip 0.5cm Heavy-flavor scale factor			  & $\pm 0.04 $\\
\hline
Method & \hspace{1ex} \\
\hskip 0.5cm Template statistics		 & $ \pm 0.18 $\\
\hskip 0.5cm Calibration			 & $ \pm 0.07 $\\
\hline
Total systematic uncertainty			& $\pm0.85$ \\
\hline
\hline
\end{tabular}
\end{table}

The systematic uncertainty due to the trigger efficiency is estimated by applying the ratio of single lepton trigger efficiency parameterization in data divided by the MC parameterization to the $ee$ and $\mu\mu$ channels.
The uncertainties in the modeling of the energy and momentum resolutions of electrons, muons, and jets
are applied independently of each other, and the shifts in \mt\ are extracted as uncertainties on \mt. Lepton energy
or momentum scales and their uncertainties are extracted from $Z\rar 2\ell$ events in data. 
An additional uncertainty is estimated 
for jet identification by shifting the jet identification efficiency within its uncertainty in 
MC samples to estimate their effect on \mt.
The uncertainty from modeling $b$ tagging is evaluated by changing within their uncertainties the corrections that account for the agreement between data and MC in $b$ tagging efficiency.

Higher-order virtual corrections to \mt\ are absent in the \alpgen\ used to generate our standard \ttbar\  
samples.  We therefore compare an ensemble of pseudo-experiments using \mcnlo\ 3.4~\cite{mcnlo} 
\ttbar\ events with one using
\alpgen\ events, where both employ \herwig\ 6.510~\cite{herwig} for modeling of hadronization. 
To evaluate the uncertainty associated with the modeling of initial and final-state radiation (ISR/FSR),
we compare \alpgen+\pythia\ with the renormalization and factorization scale
changed up and down by a factor of 1.5~\cite{ljets14}. 
The \ljets\ analysis exhibits a discrepancy in the shape of the $p_T$ distribution of the \ttbar\ system, 
which, although the dilepton statistics are limited, may be present in the dilepton sample.
We evaluate the uncertainty in the modeling of the \ttbar\ $p_T$ distribution by
reweighting MC events to make them match the data. 
The observed shift in \mt\ is taken as the uncertainty.
Since the hadronization in our standard \ttbar\ sample is
modeled with \pythia, we estimate a hadronization uncertainty on \mt\ by performing
pseudo-experiments using an \alpgen+\herwig\ sample.
We evaluate the effect of color reconnection by comparing \mt\ measurements
in \alpgen+\pythia\
samples with two \pythia\ tunes: the Perugia2011 tune that incorporates an explicit
color-reconnection scheme, and the Perugia2011 NOCR tune that does not~\cite{perugia}.
Data and MC may have different distributions in instantaneous luminosity after event selection.
This uncertainty due to multiple $p\bar{p}$ interactions is estimated by reweighting the
distribution of instantaneous luminosity to make MC agree with the data for respective data-taking epochs,
and then take the shift in \mt\ with respect to the default value.
The uncertainty due to the proton
structure is obtained from the 20 sets of CTEQ6L1 parton distribution functions (PDF) 
reweighted to CTEQ6M, where
the deviations in \mt\ for the 20 eigenvectors sets are added in quadrature~\cite{pdf}.

We estimate the effect of the uncertainty on the fraction of signal or background by changing the expected
\ttbar\ event yields ($n_{\rm S}$) up and down and the expected background yields
($n_{\rm B}$) down and up within their total uncertainties.  
The heavy-flavor scale factor, which is applied to the $Z\rar 2\ell$ cross section to correct the heavy-flavor content,
is also changed up and down within its uncertainty to estimate its systematic effect on \mt.

Our templates are constructed from MC samples for \ttbar, $Z\rar 2\ell$, and diboson backgrounds,
as well as data samples for instrumental background,
yielding statistical uncertainties on their bin contents. We use Poisson distributions to
modify bin contents within their statistical uncertainties to 
obtain 1000 new templates. We measure \mt\ in data using these templates, and the 
RMS of the measured top quark mass is taken as its uncertainty. 
Our method of \mt\ extraction relies on the correction of the fitted \mt\
to the input MC mass. The uncertainties from this calibration are applied 
to provide the uncertainty in \mt.
%All systematic uncertainties are provided in Table~\ref{tab:syst}. 
The uncertainty is reduced substantially from Ref.~\cite{llmtop12} 
due primarily to the reduction in the uncertainty in jet energy calibration and 
the optimizations for improvements in statistical uncertainty.
Larger MC samples also contribute by lowering statistical fluctuations on
systematic uncertainties, or reducing statistically limited systematic uncertainties.

\section{Conclusions}

We have measured the top quark mass in the combined dilepton channels ($e\mu$, $ee$, $\mu\mu$):
\begin{align*}
\mt &= 173.32\pm1.36({\rm stat})\pm0.85({\rm syst}){\rm  ~GeV} \\
&= 173.32\pm1.60{\rm ~GeV}.
\end{align*}
This measurement is consistent with the current average value
of \mt~\cite{worldavg}. 
Our measurement is the most precise dilepton result from the Tevatron, and is 
competitive with the most recent LHC dilepton measurements.  The systematic uncertainty of 0.49\% is
the smallest of all dilepton measurements.

\input acknowledgement.tex % input acknowledgement

\end{document}

%% file: author_list.tex
\affiliation{LAFEX, Centro Brasileiro de Pesquisas F\'{i}sicas, Rio de Janeiro, Brazil}
\affiliation{Universidade do Estado do Rio de Janeiro, Rio de Janeiro, Brazil}
\affiliation{Universidade Federal do ABC, Santo Andr\'e, Brazil}
\affiliation{University of Science and Technology of China, Hefei, People's Republic of China}
\affiliation{Universidad de los Andes, Bogot\'a, Colombia}
\affiliation{Charles University, Faculty of Mathematics and Physics, Center for Particle Physics, Prague, Czech Republic}
\affiliation{Czech Technical University in Prague, Prague, Czech Republic}
\affiliation{Institute of Physics, Academy of Sciences of the Czech Republic, Prague, Czech Republic}
\affiliation{Universidad San Francisco de Quito, Quito, Ecuador}
\affiliation{LPC, Universit\'e Blaise Pascal, CNRS/IN2P3, Clermont, France}
\affiliation{LPSC, Universit\'e Joseph Fourier Grenoble 1, CNRS/IN2P3, Institut National Polytechnique de Grenoble, Grenoble, France}
\affiliation{CPPM, Aix-Marseille Universit\'e, CNRS/IN2P3, Marseille, France}
\affiliation{LAL, Universit\'e Paris-Sud, CNRS/IN2P3, Orsay, France}
\affiliation{LPNHE, Universit\'es Paris VI and VII, CNRS/IN2P3, Paris, France}
\affiliation{CEA, Irfu, SPP, Saclay, France}
\affiliation{IPHC, Universit\'e de Strasbourg, CNRS/IN2P3, Strasbourg, France}
\affiliation{IPNL, Universit\'e Lyon 1, CNRS/IN2P3, Villeurbanne, France and Universit\'e de Lyon, Lyon, France}
\affiliation{III. Physikalisches Institut A, RWTH Aachen University, Aachen, Germany}
\affiliation{Physikalisches Institut, Universit\"at Freiburg, Freiburg, Germany}
\affiliation{II. Physikalisches Institut, Georg-August-Universit\"at G\"ottingen, G\"ottingen, Germany}
\affiliation{Institut f\"ur Physik, Universit\"at Mainz, Mainz, Germany}
\affiliation{Ludwig-Maximilians-Universit\"at M\"unchen, M\"unchen, Germany}
\affiliation{Panjab University, Chandigarh, India}
\affiliation{Delhi University, Delhi, India}
\affiliation{Tata Institute of Fundamental Research, Mumbai, India}
\affiliation{University College Dublin, Dublin, Ireland}
\affiliation{Korea Detector Laboratory, Korea University, Seoul, Korea}
\affiliation{CINVESTAV, Mexico City, Mexico}
\affiliation{Nikhef, Science Park, Amsterdam, the Netherlands}
\affiliation{Radboud University Nijmegen, Nijmegen, the Netherlands}
\affiliation{Joint Institute for Nuclear Research, Dubna, Russia}
\affiliation{Institute for Theoretical and Experimental Physics, Moscow, Russia}
\affiliation{Moscow State University, Moscow, Russia}
\affiliation{Institute for High Energy Physics, Protvino, Russia}
\affiliation{Petersburg Nuclear Physics Institute, St. Petersburg, Russia}
\affiliation{Instituci\'{o} Catalana de Recerca i Estudis Avan\c{c}ats (ICREA) and Institut de F\'{i}sica d'Altes Energies (IFAE), Barcelona, Spain}
\affiliation{Uppsala University, Uppsala, Sweden}
\affiliation{Taras Shevchenko National University of Kyiv, Kiev, Ukraine}
\affiliation{Lancaster University, Lancaster LA1 4YB, United Kingdom}
\affiliation{Imperial College London, London SW7 2AZ, United Kingdom}
\affiliation{The University of Manchester, Manchester M13 9PL, United Kingdom}
\affiliation{University of Arizona, Tucson, Arizona 85721, USA}
\affiliation{University of California Riverside, Riverside, California 92521, USA}
\affiliation{Florida State University, Tallahassee, Florida 32306, USA}
\affiliation{Fermi National Accelerator Laboratory, Batavia, Illinois 60510, USA}
\affiliation{University of Illinois at Chicago, Chicago, Illinois 60607, USA}
\affiliation{Northern Illinois University, DeKalb, Illinois 60115, USA}
\affiliation{Northwestern University, Evanston, Illinois 60208, USA}
\affiliation{Indiana University, Bloomington, Indiana 47405, USA}
\affiliation{Purdue University Calumet, Hammond, Indiana 46323, USA}
\affiliation{University of Notre Dame, Notre Dame, Indiana 46556, USA}
\affiliation{Iowa State University, Ames, Iowa 50011, USA}
\affiliation{University of Kansas, Lawrence, Kansas 66045, USA}
\affiliation{Louisiana Tech University, Ruston, Louisiana 71272, USA}
\affiliation{Northeastern University, Boston, Massachusetts 02115, USA}
\affiliation{University of Michigan, Ann Arbor, Michigan 48109, USA}
\affiliation{Michigan State University, East Lansing, Michigan 48824, USA}
\affiliation{University of Mississippi, University, Mississippi 38677, USA}
\affiliation{University of Nebraska, Lincoln, Nebraska 68588, USA}
\affiliation{Rutgers University, Piscataway, New Jersey 08855, USA}
\affiliation{Princeton University, Princeton, New Jersey 08544, USA}
\affiliation{State University of New York, Buffalo, New York 14260, USA}
\affiliation{University of Rochester, Rochester, New York 14627, USA}
\affiliation{State University of New York, Stony Brook, New York 11794, USA}
\affiliation{Brookhaven National Laboratory, Upton, New York 11973, USA}
\affiliation{Langston University, Langston, Oklahoma 73050, USA}
\affiliation{University of Oklahoma, Norman, Oklahoma 73019, USA}
\affiliation{Oklahoma State University, Stillwater, Oklahoma 74078, USA}
\affiliation{Brown University, Providence, Rhode Island 02912, USA}
\affiliation{University of Texas, Arlington, Texas 76019, USA}
\affiliation{Southern Methodist University, Dallas, Texas 75275, USA}
\affiliation{Rice University, Houston, Texas 77005, USA}
\affiliation{University of Virginia, Charlottesville, Virginia 22904, USA}
\affiliation{University of Washington, Seattle, Washington 98195, USA}
\author{V.M.~Abazov} \affiliation{Joint Institute for Nuclear Research, Dubna, Russia}
\author{B.~Abbott} \affiliation{University of Oklahoma, Norman, Oklahoma 73019, USA}
\author{B.S.~Acharya} \affiliation{Tata Institute of Fundamental Research, Mumbai, India}
\author{M.~Adams} \affiliation{University of Illinois at Chicago, Chicago, Illinois 60607, USA}
\author{T.~Adams} \affiliation{Florida State University, Tallahassee, Florida 32306, USA}
\author{J.P.~Agnew} \affiliation{The University of Manchester, Manchester M13 9PL, United Kingdom}
\author{G.D.~Alexeev} \affiliation{Joint Institute for Nuclear Research, Dubna, Russia}
\author{G.~Alkhazov} \affiliation{Petersburg Nuclear Physics Institute, St. Petersburg, Russia}
\author{A.~Alton$^{a}$} \affiliation{University of Michigan, Ann Arbor, Michigan 48109, USA}
\author{A.~Askew} \affiliation{Florida State University, Tallahassee, Florida 32306, USA}
% ...
%
% visitor_addresses.tex                       1 April 2015
%  available symbols are:
%  $\ast, \dag, \ddag, \S, \P, $\|$, $\ast\ast$, \dag\dag, \ddag\ddag ,\#
%
\collaboration{The D0 Collaboration\footnote{with visitors from
%{alton}
$^{a}$Augustana College, Sioux Falls, SD, USA,
%{burdin}
$^{b}$The University of Liverpool, Liverpool, UK,
%{grohsjean,deterre}
$^{c}$DESY, Hamburg, Germany,
%{de la cruz-burelo}
$^{d}$CONACyT, Mexico City, Mexico,
%{partridge}
$^{e}$SLAC, Menlo Park, CA, USA,
%{hesketh}
$^{f}$University College London, London, UK,
%{luna-garcia}
$^{g}$Centro de Investigacion en Computacion - IPN, Mexico City, Mexico,
%{santos}
$^{h}$Universidade Estadual Paulista, S\~ao Paulo, Brazil,
%{meyer}
$^{i}$Karlsruher Institut f\"ur Technologie (KIT) - Steinbuch Centre for Computing (SCC),
D-76128 Karlsruhe, Germany,
%{patwa}
$^{j}$Office of Science, U.S. Department of Energy, Washington, D.C. 20585, USA,
%{cooke}
$^{k}$American Association for the Advancement of Science, Washington, D.C. 20005, USA,
%{borysova}
$^{l}$Kiev Institute for Nuclear Research, Kiev, Ukraine,
%{jabeen}
$^{m}$University of Maryland, College Park, Maryland 20742, USA
and
%{williams}
$^{n}$European Orgnaization for Nuclear Research (CERN), Geneva, Switzerland
%{montgomery}
%$^{?}$Thomas Jefferson National Accelerator Facility, Newport News, VA 23606, USA,
%{falkowski}
%$^{?}$Laboratoire de Physique Theorique, Orsay, FR,
%{hooper,kozminski}
%$^{?}$}Visitor from Lewis University, Romeoville, IL, USA.
%{weber}
%$^{?}$Universit{\"a}t Bern, Bern, Switzerland.
%{deceased}
%{zanabria}
%$^{?}$City Colleges of Chicago, Chicago, IL, USA}
%$^{\ddag}$Deceased.
}} \noaffiliation
\vskip 0.25cm

%% file: acknowledgement.tex
We thank the staffs at Fermilab and collaborating institutions,
and acknowledge support from the
Department of Energy and National Science Foundation (United States of America);
Alternative Energies and Atomic Energy Commission and
National Center for Scientific Research/National Institute of Nuclear and Particle Physics  (France);
Ministry of Education and Science of the Russian Federation, 
National Research Center ``Kurchatov Institute" of the Russian Federation, and 
Russian Foundation for Basic Research  (Russia);
National Council for the Development of Science and Technology and
Carlos Chagas Filho Foundation for the Support of Research in the State of Rio de Janeiro (Brazil);
Department of Atomic Energy and Department of Science and Technology (India);
Administrative Department of Science, Technology and Innovation (Colombia);
National Council of Science and Technology (Mexico);
National Research Foundation of Korea (Korea);
Foundation for Fundamental Research on Matter (The Netherlands);
Science and Technology Facilities Council and The Royal Society (United Kingdom);
Ministry of Education, Youth and Sports (Czech Republic);
Bundesministerium f\"{u}r Bildung und Forschung (Federal Ministry of Education and Research) and 
Deutsche Forschungsgemeinschaft (German Research Foundation) (Germany);
Science Foundation Ireland (Ireland);
Swedish Research Council (Sweden);
China Academy of Sciences and National Natural Science Foundation of China (China);
and
Ministry of Education and Science of Ukraine (Ukraine).

%% file: llmtop.bbl
\begin{thebibliography}{99}
\bibitem{pseudorapidity} The pseudorapidity is defined as $\eta = - \ln[\tan(\theta /2)]$ where $\theta$ is the polar angle relative to the proton beam direction.
\end{thebibliography}
